\documentclass[12pt,preprint]{aastex}

\newcommand{\bm}[1]{\mbox{\boldmath{$#1$}}}

\newcommand{\at}{{\rm Athena}}
\newcommand{\del}{{\bf \nabla}}

\newcommand{\md}{\dot{M}}
\newcommand{\kd}{\dot{K}}

\newcommand{\gd}{\dot{G}}
\newcommand{\qk}{Q_{\rm k}}
\newcommand{\qm}{Q_{\rm m}}

\newcommand{\alf}{{\rm Alfv\acute{e}n}}
\newcommand{\pr}{P_{\rm m}}
\newcommand{\cs}{c_{\rm s}}
\newcommand{\va}{v_{\rm A}}

\begin{document}

\title{Viscous and Resistive Effects on the MRI with a Net Toroidal Field}

\author{Jacob B. Simon, John F. Hawley}
\affil{Department of Astronomy \\ University of Virginia \\ P.O. Box 400325 \\
Charlottesville, VA 22904-4325}

\begin{abstract}

Resistivity and viscosity have a significant
role in establishing the energy levels in turbulence driven by the
magnetorotational instability (MRI) in local astrophysical disk models.
This study uses the $\at$ code to characterize the effects of a constant
shear viscosity $\nu$ and Ohmic resistivity $\eta$ in unstratified
shearing box simulations with a net toroidal magnetic flux.  
A previous study of shearing boxes
with zero net magnetic field performed with the ZEUS code found that
turbulence dies out for values of the magnetic Prandtl number, $\pr = \nu/\eta$, 
below $\pr \sim 1$; for $\pr \gtrsim 1$, time- and 
volume-averaged stress levels increase with $\pr$. We repeat these experiments with $\at$ and obtain
consistent results.  Next, the influence of viscosity and resistivity
on the toroidal field MRI is investigated both for linear growth and
for fully-developed turbulence.  In the linear regime, a sufficiently
large $\nu$ or $\eta$ can prevent MRI growth; $\pr$ itself has
little direct influence on growth from linear perturbations.  By applying
a range of values for $\nu$ and $\eta$ to an initial state consisting
of fully developed turbulence in the presence of a background toroidal
field, we investigate their effects in the fully nonlinear system.  Here,
increased viscosity enhances the turbulence, and the turbulence decays
only if the resistivity is above a critical value; turbulence can be
sustained even when $\pr < 1$, in contrast to the zero net field model.
While we find preliminary evidence that the stress converges to a small range
of values when $\nu$ and $\eta$ become small enough,
the influence of dissipation terms on MRI-driven turbulence for relatively
large $\eta$ and $\nu$ is significant, independent of field geometry.

\end{abstract}


\section{Introduction}
\label{introduction}

Disk accretion is a fundamental process of many astrophysical phenomena,
from nearby young stellar objects to immensely luminous distant quasars.
Understanding the mechanism for removing angular momentum from a
fluid element, thereby allowing accretion to occur, is essential to
understanding these systems.  Orbiting, magnetized gas is unstable to the
magnetorotational instability (MRI) \cite[]{balbus91,balbus98};
all that is required is a subthermal magnetic field sufficiently coupled
to differentially rotating gas with a negative outward angular velocity
gradient.  The MRI leads to turbulent flow, resulting in Maxwell and
Reynolds stresses that efficiently transport angular momentum and drive
accretion.  However, it is still uncertain what determines the amplitude
of the magnetic energy and stress in saturated MRI-driven turbulence.

Because linear analysis can offer only limited guidance, numerical
simulations have been used to investigate the properties of MRI-driven
turbulence.  Most simulations of the MRI employ the shearing box
approximation, in which the simulation domain consists of a local
corotating patch of accretion disk, small enough to expand the MHD
equations into Cartesian coordinates and ignore curvature terms
\cite[see][]{hawley95a}.  This approximation, in its simplest form,
reduces the problem to its basic ingredients:  differential rotation
and magnetized fluid.  It is hoped that a more complete understanding
of this simple model will provide insights into the mechanisms that
determine the stress levels in astrophysical systems.

The first shearing box simulations \citep{hawley95a,hawley96} found
that the presence of a net magnetic field and its orientation play a
role in setting the amplitude of the MRI turbulence.  A net magnetic
field results from currents located outside of the computational domain
and cannot change as a result of the evolution.  For zero net field
simulations, on the other hand, complete decay of the field is possible.
Net vertical fields gave the largest turbulent energies, with the
energy level approximately proportional to the background field.
Net toroidal fields behave similarly, but with a smaller energy for the
same background field strength.  Zero net field simulations saturated
at levels comparable to those seen in net toroidal field cases.

In subsequent years, there have been many shearing box simulations which
confirm these qualitative behaviors, but the factors that determine
the amplitude of the turbulent energy still remain uncertain.  Some studies
have examined numerical effects such as computational domain size
and resolution, and others have looked at physical parameters such as
background field strength and gas pressure.  An investigation
of the influence of gas pressure carried out by \cite{sano04}, for
example, found an extremely weak pressure dependence. Even here, the
influence of the gas pressure depends on the magnetic field geometry.
\cite{blackman08} examined the results of an ensemble of shearing box
simulations taken from the literature and found that $\alpha\beta$ is
generally constant, where $\alpha$ is the total stress divided by the
gas pressure, and $\beta$ is the ratio of thermal to magnetic pressure.
In other words, stress is proportional to magnetic energy.

A physical influence that has, until recently, received less
attention is physical dissipation, namely shear viscosity $\nu$ and Ohmic
resistivity $\eta$.  The linear dispersion relation for the vertical field
MRI in the presence of $\nu$ and $\eta$ was derived by \cite{balbus98}.
Both terms can reduce the effectiveness of the MRI.  In the linear regime,
viscosity damps the MRI growth rates and changes the wavelength of the
fastest growing mode, but does not alter the wavenumbers that are unstable.
Resistivity introduces a cutoff on the unstable wavelengths where the
resistive diffusion time becomes comparable to the MRI growth time
\citep[see, e.g., the discussion in][]{masada08}.  Nonaxisymmetric MRI
modes with Ohmic resistivity were examined by \cite{papaloizou97}.
They found that resistivity reduces the amplification of such modes,
and if large enough, can stabilize the toroidal field MRI.

Simulations by \cite{hawley96}, \cite{sano98},
\cite{fleming00}, \cite{sano01}, \cite{ziegler01}, and \cite{sano02b}
have investigated the influence of a nonzero Ohmic resistivity on the
saturated state.  The main result of these studies is that increasing
the resistivity leads to a decrease in turbulence, independent of the
magnetic field configuration.  In zero net field models, the effect of
resistivity on the turbulence is larger than one might expect from the
linear MRI relation \cite[]{fleming00}, with the turbulence decaying to
zero for relatively low values of resistivity.

Recently, the work of \cite{fromang07b} (hereafter F07) and \cite{lesur07}
has sparked new interest in the effects of non-ideal MHD on the MRI.
F07 showed that both resistivity and viscosity are important in
determining
the stress level in MRI turbulent flows with zero net magnetic field.
\cite{lesur07} came to the same conclusion for models with a net vertical
field.  The results were characterized in terms of the magnetic Prandtl
number, defined as $\pr = \nu/\eta$.  In these simulations, the saturation
level increases with increasing $\pr$.  F07 also find that for the zero
net field case, there exists a $\pr$ below which the turbulence dies
out, and that this critical $\pr$ decreases with decreasing
viscosity (at least for the range in viscosity and resistivity examined
in the paper).

One magnetic field geometry that has not yet been explored
with both physical resistivity and viscosity is that of a net toroidal
field.  Such fields could be the most relevant to astrophysical disks.
Following the arguments of \cite{guan09a} and references therein, both
global and local disk simulations as well as observations of disk
galaxies show a dominance of toroidal field over other field components.
Indeed, the background shear flow naturally creates toroidal field from
radial field. It seems likely that any given region of an accretion disk
will contain some net azimuthal field.

In this paper, we perform the first investigation of the toroidal
field MRI in the presence of {\it both} viscosity and resistivity
and compare the outcomes with those obtained for zero net and net
vertical field simulations.  The structure of the paper is as follows.
In \S~\ref{method}, we describe our algorithm, parameters, and tests
of our viscosity and resistivity implementation. For comparison purposes,
we reexamine the simulations of F07 with our code in
\S~\ref{zero_net_flux}.  Our main results, focusing on the toroidal
field simulations, are presented in \S~\ref{toroidal_field}.  We wrap
up with our discussion and conclusions in \S~\ref{conclusions}.

\section{Numerical Simulations}
\label{method}

In this study, we use the $\at$ code, a second-order accurate Godunov
scheme for solving the equations of MHD in conservative form using
the dimensionally unsplit corner transport upwind (CTU) method of
\cite{colella90} coupled with the third-order in space piecewise
parabolic method (PPM) of \cite{colella84} and a constrained transport
\citep[CT;][]{evans88} algorithm for preserving the $\del \cdot {\bm
B}$~=~0 constraint.  A description of this algorithm and various
test problems is given in \cite{gardiner05a}, \cite{gardiner08},
and \cite{stone08}.  For the present study, we have added physical
dissipation in the form of a constant kinematic shear viscosity and Ohmic
resistivity using operator splitting, as described in more detail below.
Bulk viscosity is ignored.

The shearing box approximation is a model for a local region of a disk
orbiting at a radius $R$ whose size is small compared to this radius,
allowing us to expand the equations of motion in Cartesian form, as
described in detail by \cite{hawley95a}.  The box corotates with an
angular velocity $\Omega$ corresponding to the value at the center of
the box.  The shearing box evolution equations with viscosity and
resistivity are given by \cite{balbus98} and are:
\begin{equation}
\label{cont}
\frac{\partial \rho}{\partial t} + \del \cdot (\rho {\bm v}) = 0,
\end{equation}
\begin{equation}
\label{momentum}
\frac{\partial \rho {\bm v}}{\partial t} + \del \cdot (\rho {\bm v}{\bm v} - {\bm B}{\bm B}) + 
\del \left(P + \frac{1}{2} B^2\right) = 
2 q \rho \Omega^2 {\bm x} - 2 {\bm \Omega} \times \rho {\bm v} + \del \cdot (\rho \nu \del {\bm v}) + \del\left(\frac{1}{3}\rho\nu\del \cdot {\bm v}\right),
\end{equation}
\begin{equation}
\label{induction}
\frac{\partial {\bm B}}{\partial t} = \del \times ({\bm v} \times {\bm B} - \eta \del \times {\bm B}),
\end{equation}

\noindent 
where $\rho$ is the mass density, $\rho {\bm v}$ is the momentum
density, ${\bm B}$ is the magnetic field, $P$ is the gas pressure, and $q$
is the shear parameter, defined as $q = -d$ln$\Omega/d$ln$R$.  We use $q =
3/2$, appropriate for a Keplerian disk.  We assume an isothermal equation
of state $P = \rho \cs^2$, where $\cs$ is the isothermal sound speed.
Shear viscosity and Ohmic resistivity are denoted by $\nu$ and $\eta$
respectively.  Note that our system of units has the magnetic permeability
$\mu = 1$.  The first source term on the right-hand side
of equation~(\ref{momentum}) corresponds to tidal forces (gravity
and centrifugal) in the corotating frame.  The second source term in
equation~(\ref{momentum}) is the Coriolis force.  
Finally, we have omitted the vertical component of gravity, making
these ``unstratified'' shearing box simulations.

Adapting the $\at$ code to the shearing box problem requires adding the
tidal and Coriolis force terms and implementing the shearing-periodic
boundary conditions at the $x$ boundaries.  The source terms are included
in the algorithm in a directionally unsplit manner, consistent with the
CTU algorithm.  We do not use the Crank-Nicholson method of
\cite{gardiner05b} that ensures precise conservation of epicyclic
energy. We have found this added complexity to be 
unnecessary for simulations dominated by the MRI
\cite[see arguments in][]{simon09}.  The shearing-periodic boundary
conditions are described in \cite{simon09}.  Quantities are linearly
reconstructed in the ghost zones from appropriate zones in the physical
domain that have been shifted along $y$ to account for the shear across
the boundaries.  Furthermore, the $y$ momentum is adjusted to account for
the shear across the $x$ boundaries as fluid moves out one boundary and
enters at the other.

Note that to preserve a quantity to machine precision across a
grid boundary such as the shearing-periodic boundary (or a boundary
between different grids in a mesh-refinement scheme), it is necessary
to reconstruct a quantity's flux (or for the magnetic field, the
electromotive force, EMF) at the boundary, rather than the quantity
itself \cite[see][]{simon09}.  To conserve total vertical field flux,
for example, we reconstruct the $y$ EMF at the $x$ boundaries.  This is
essential, given the strong effect that a net vertical field has on
the turbulence level.  The perfect conservation of net toroidal flux is
not as important, and as ensuring its precise conservation involves a
more complex procedure, we allow the $B_y$ flux to be conserved only to
truncation level.  Note that in our simulations initialized with a net toroidal field, this truncation error
results in a loss of net $B_y$ flux from the domain; $\sim$ 5-10 \% of the
initial toroidal field is lost per 100 orbits for our high resolution, sustained turbulence simulations.
This corresponds to a background $\beta$ value of $\sim$ 110-120 at 100 orbits.
While this truncation error does not appear to have any significant affect on
the turbulent energy levels in our simulations, it may become important to conserve $B_y$ to roundoff
level for longer evolution times.
 The radial flux, $B_x$, will automatically be conserved
to machine precision because its evolution is determined by EMFs on
the periodic $y$ and $z$ boundaries.  We also reconstruct the density
flux on the $x$ boundaries to conserve the total mass in the domain to
machine precision.  The systematic difference between the calculation of
outward and inward fluxes at the shearing $x$ boundaries can lead to a loss
of mass from the grid \citep[]{simon09}. We do not reconstruct momentum fluxes at the boundaries as
the source terms will prevent roundoff level conservation of momentum.

Both the viscosity and resistivity are added via operator splitting;
the fluid variables updated from the CTU integrator are used to
calculate the viscous and resistive terms on the right-hand side of
equations~(\ref{momentum}) and (\ref{induction}).  These terms are
discretized in a flux-conservative manner consistent with the $\at$
algorithm.  In particular, the third and fourth terms on the right-hand
side of equation~(\ref{momentum}) are written so that $\rho \nu \del {\bm
v}$ and $(1/3)\rho\nu \del \cdot {\bm v}$ are defined as fluxes at the
cell faces.  Taking the divergence of the third term and the gradient of
the fourth term via finite-differencing ensures that momentum conservation
is not violated by the viscous terms.  The resistive contribution to the
induction equation is added in a manner consistent with the EMFs; the term
$\eta \del \times {\bm B}$ is computed at cell corners to ensure that when
differenced via the curl operator, $\del \cdot {\bm B} = 0$ is maintained.
Note that this resistive contribution to the EMF must also be reconstructed
at the shearing-periodic boundaries in order to preserve $B_z$ precisely.

More generally, the viscous term in equation~(\ref{momentum})
can be written in a flux-conservative manner as $\del \cdot {\bm T}$
where ${\bm T}$ is a viscous stress tensor defined as
\begin{equation}
\label{stress_tensor}
T_{ij} = \rho \nu \left(\frac{\partial v_i}{\partial x_j} + \frac{\partial v_j}{\partial x_i} - \frac{2}{3}\delta_{ij} \del \cdot {\bm v} \right),
\end{equation}
\noindent where the indices refer to the spatial components \cite[]{landau59}.
For simplicity, we have used
the form as in equation~(\ref{momentum})  which is equivalent to
equation~(\ref{stress_tensor}) assuming that $\rho \nu$ is spatially
constant.  We have performed a few shearing box experiments with both
implementations, and find no significant differences in turbulent
stress evolution.  In particular, we restarted a few simulations
using the form in equation~(\ref{stress_tensor}).  We found that the
volume-averaged magnetic energies are initially indistinguishable between
the two approaches.  Due to the chaotic nature of the MRI, the two curves
eventually diverge, but nevertheless maintain the same time average.

The addition of viscosity and resistivity places an
additional constraint on the time step,
\begin{equation}
\label{timestep}
\Delta t = C_o {\rm MIN}\left(\Delta t_{\rm CTU}, 0.75 \frac{\Delta^2}{8/3 \nu}, 0.75 \frac{\Delta^2}{2 \eta}\right),
\end{equation}

\noindent where $C_o$ is the CFL number ($C_o$ = 0.4 here), $\Delta
t_{\rm CTU}$ is the time step limit from the main integration algorithm
\cite[see][]{stone08}, and $\Delta$ is the minimum grid spacing, $\Delta
= {\rm MIN}(\Delta x, \Delta y, \Delta z)$.  Several three-dimensional
tests of viscosity and resistivity revealed that if the viscous or
resistive time step is close to $\Delta t_{\rm CTU}$, the evolution
becomes numerically unstable.  This problem was remedied by multiplying
the viscous and resistive time steps by 0.75.   The additional 4/3 factor
in the denominator of the viscous time step results from the last term
on the right-hand side of equation~(\ref{momentum}).  This can be most
easily understood by considering a one-dimensional problem, in which
case the effective $\nu$ value increases by a factor of 4/3 due to the
compressibility term.  Therefore, the effective $\nu$ that goes into
the time step calculation is taken as $(4/3) \nu$.  Note that most of
our simulations will have $\nu$ and $\eta$ sufficiently small that the
viscous and resistive time steps are large compared to $\Delta t_{\rm
CTU}$. In fact, only the simulations with the largest values of $\eta$
and $\nu$ reach the diffusion limit on $\Delta t$.

\subsection{Tests of Physical Dissipation}
\label{diss_test}

We performed a number of problems to test the implementation of
viscosity and resistivity within $\at$.  
Resistivity was tested by solving the
diffusion of a current sheet along one dimension; a uniform magnetic
field is initialized with a change in sign across one grid zone. This
problem has a simple analytic solution \cite[see e.g.,][]{komissarov07}.
The agreement between the numerical and analytic solution was excellent.
By replacing the magnetic field with a uniform velocity flow, the identical
test can be performed for the viscosity.  Again, the numerical solution
agreed with the analytic solution.

Next, we initialized a uniform vertical magnetic field in
a shearing box with nonzero viscosity and resistivity and measured the
growth of various MRI modes in the linear regime. We compared the measured
values with those from analytic linear theory \cite[see e.g.,][]{masada08, pessah08} and
found excellent agreement for a wide range in viscosity and resistivity.

\begin{figure}
\plotfiddle{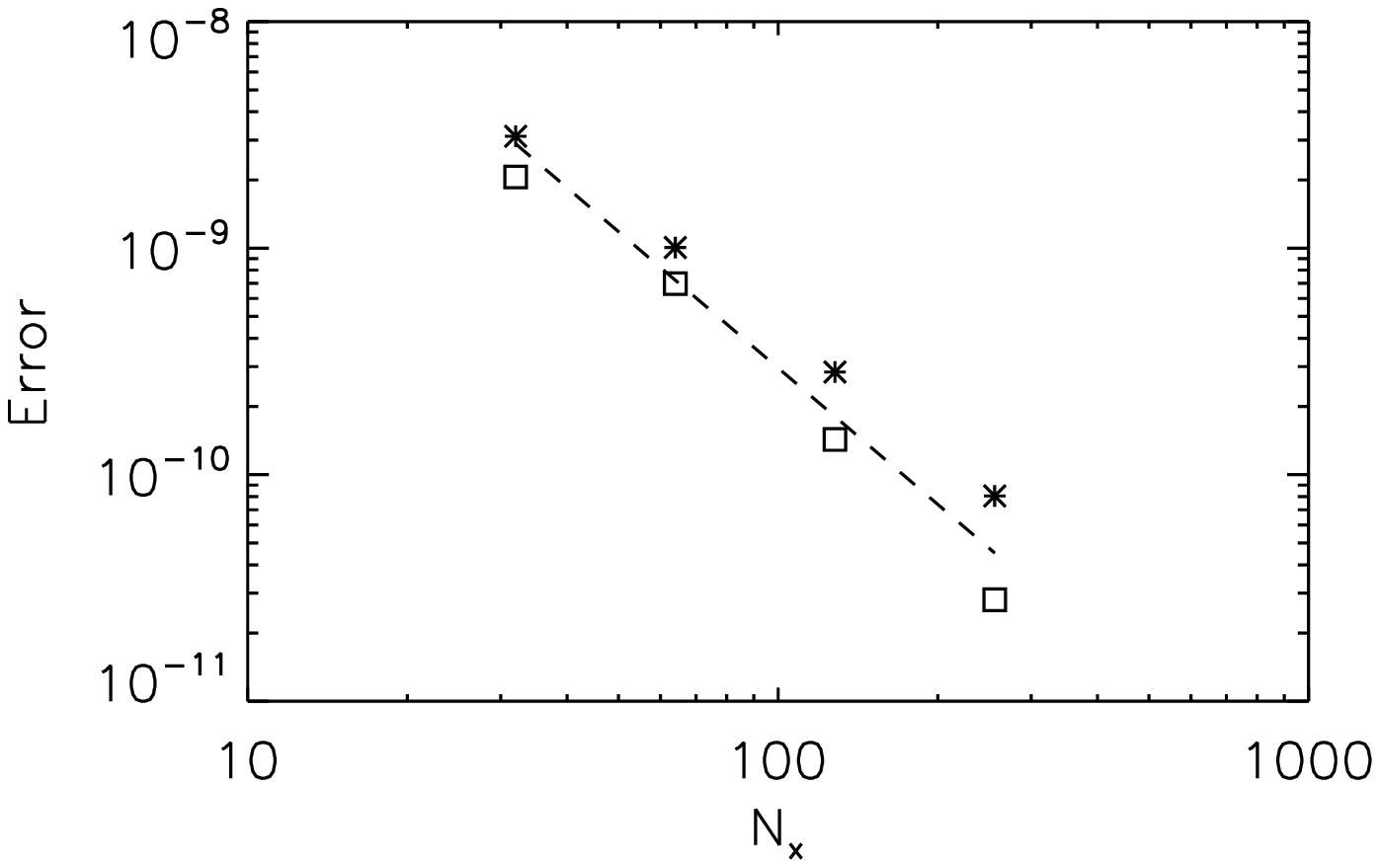}{0in}{0.}{400}{270}{10}{-100}
\vspace{-0.2in}
\caption{Numerical error as a
function of $x$ resolution for the three-dimensional decaying
linear wave problem.  The boxes are the errors for
a decaying $\alf$ wave, and the asterisks are the errors for a decaying
sound wave.  The error is calculated from the square root of the sum
of the squared errors in the density and momenta (for the sound wave)
and the density, momenta, and magnetic field (for the $\alf$ wave)
obtained using the analytic solution.  The
dashed line shows the slope corresponding to 
second-order convergence.
}
\label{lw_conv}
\end{figure}

Finally, we examined the propagation of small amplitude, isothermal
sound and $\alf$ waves in the presence of viscosity and resistivity.
Again, the numerical solution can be compared directly to an analytic
solution.  These tests were done in one, two, and three dimensions;
in the multidimensional tests, the propagation direction of the wave was
chosen to be along the grid diagonal.  The resistivity was tested via
the decay of the $\alf$ waves, and the viscosity was tested via the
decay of the sound waves. The error as a function of $x$ resolution
for two of these tests is given in Fig.~\ref{lw_conv}.  The error
is calculated from the square root of the sum of the squared errors in
the density and momenta (for the sound wave) and the density, momenta,
and magnetic field (for the $\alf$ wave).  The solution to each wave
converges at a rate very close to second order, shown by the dashed line.

\subsection{Shearing Box Parameters}
\label{init_conds}

The shearing box used in this study has radial size $L_x = 1$, azimuthal
size $L_y = 4$, and vertical size $L_z = 1$.  Most of the simulations
presented here use $128\times 200 \times 128$ equally spaced grid zones;
some simulations use half the number of zones in each direction.  The
initial velocity is ${\bm v} = -q\Omega x \hat{{\bm y}}$, with $q$~=~3/2,
$\Omega$~=~0.001, and $-L_x/2 \leq x \leq L_x/2$.  The isothermal sound
speed is $\cs = \Omega H$ where $H$ is the scale height.  With $L_z =
H$, we have $\cs$~=~$L_z\Omega$, and with $\rho=1$, the initial pressure
is $P = \rho \Omega^2 L_z^2 = 10^{-6}$.

The dissipation terms $\nu$ and $\eta$ are parameterized in terms of
the Reynolds number,

\begin{equation}
\label{reynolds}
Re \equiv \frac{\cs H }{\nu},
\end{equation}

\noindent
the magnetic Reynolds number,

\begin{equation}
\label{mag_reynolds}
Rm \equiv \frac{\cs H}{\eta},
\end{equation}

\noindent
and the magnetic Prandtl number,

\begin{equation}
\label{prandtl}
\pr \equiv \frac{\nu}{\eta} = \frac{Rm}{Re}.
\end{equation}

\noindent Since the properties of the MRI are more directly
determined by the $\alf$ speed rather than the sound speed, another useful
dimensionless quantity is the Elsasser number,

\begin{equation}
\label{elsasser}
\Lambda \equiv \frac{\va^2}{\eta \Omega},
\end{equation}

\noindent where $\va$ is the $\alf$ speed.  With $\cs = \Omega H$ and
$\beta = 2 \cs^2/\va^2$, we can relate $Rm$ to $\Lambda$,

\begin{equation}
\label{rm_els_relate}
\Lambda = \frac{2}{\beta} Rm.
\end{equation}

In addition to the explicit dissipation terms, there will also be some
effective diffusion due to numerical effects.  Generally speaking,
numerical diffusion will not behave in the same manner as physical
diffusion (e.g., it is not a simple function of a gradient in field or
velocity); numerical diffusion generally has a much stronger effect at
small scales than at large scales.  Also the effects of numerical diffusion
may be different from one type of simulation to another.  By calculating
numerical losses at high wavenmbers in Fourier space and modeling those
as if they were physical viscosity and resistivity, \cite{simon09}
quantified the numerical dissipation of $\at$.  They found that the
effective $Rm$ for the zero net field and net z field simulations at
$N_x = 128$ were 20000 and 8000 respectively, and 7000 and 5000 for
$N_x = 64$.  The effective $\pr$ is $\sim$ 2 for these simulations.
Since numerical dissipation is problem-dependent, these numbers should
be regarded as estimates, and their values will likely be somewhat
different in different applications.  However, they serve as a guideline
for including physical dissipation.  In the present study, numerical
and physical dissipation may be comparable at large wavenumbers for $Re,
Rm \gtrsim 10000$.  The physical dissipation in some of our simulations
may fall into this marginally resolved regime.  Nevertheless, we can
explore a large enough range in $Re$ and $Rm$ values to observe clear
effects due to viscosity and resistivity.

\section{Zero Net Flux Simulations}
\label{zero_net_flux}

\cite{fromang07a} and \cite{pessah07} presented the surprising result that for zero net
field shearing box simulations without any explicit dissipation terms, 
the steady-state turbulent energy decreases  with increasing grid resolution. 
\cite{simon09} obtained the same result for zero net field simulations
without explicit dissipation using the $\at$ code.  These results pointed 
to the importance of including explicit dissipation terms in such simulations.

F07 showed that turbulent activity is strongly influenced
by these dissipation terms; the saturated stress increases with
increasing $\pr$.  Here we return to the zero net
field problem and include the dissipative terms to compare with the
results of F07.   The simulations are initialized with ${\bm B} =
\sqrt{2P/\beta}\,{\rm sin[}(2\pi/L_x) x{\rm]}\hat{{\bm z}}$ where
$\beta$ = 400. These runs are labeled SZ for sinusoidal z-field and
have resolution $N_x$~=~128, $N_y$~=~200, $N_z$~=~128. The viscosity and
resistivity in these simulations are chosen to reproduce the calculations
of F07.   The initial state is perturbed in each grid zone with random
fluctuations in $\rho$ at amplitude $\delta \rho/\rho$~=~0.01.  The SZ
simulations are listed in Table~\ref{tbl:sz_runs}.  The column labeled
``Turbulence?"  states whether or not the turbulence was sustained in a
given simulation.  The column labeled ``$\alpha$" gives the resulting
turbulent stress in terms of the dimensionless value $\alpha \equiv
\langle\langle \rho v_x\delta v_y - B_xB_y \rangle\rangle/P_o$, with 
$\delta v_y \equiv v_y + q\Omega x$.  $P_o$
is the initial gas pressure and the double bracket denotes a time and
volume average. The time average is calculated from orbit 20 until the end of the simulation,
and as is the case throughout this paper, volume average refers to an average
over the entire simulation domain.

\begin{figure}
\plotfiddle{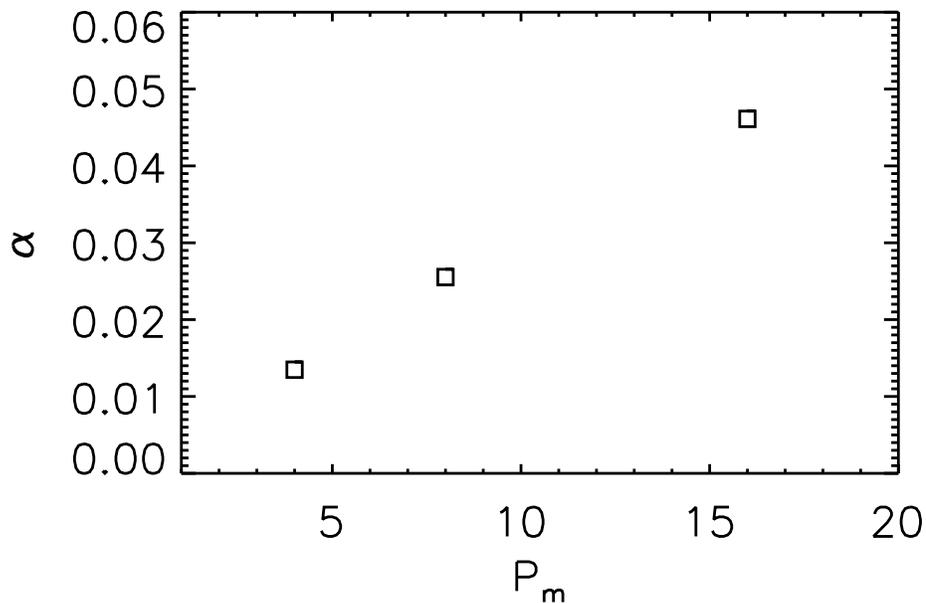}{0in}{0.}{400}{270}{10}{-100}
\vspace{-0.2in}
\caption{Time- and volume-averaged stress parameter $\alpha$ as a
function of $\pr$ in the SZ simulations; $\alpha \equiv \langle\langle
\rho v_x\delta v_y - B_xB_y \rangle\rangle/P_o$, where the average is calculated over the entire
simulation domain and from 20 orbits to the end of the simulation.  Only simulations with sustained
turbulence are plotted.  The $\pr = 4$ model has $Rm = 12500$ whereas
the other two have $Rm = 12800$.  There is a nearly linear increase in
$\alpha$ with $\pr$.
}
\label{sz_rm12800}
\end{figure}

The results of these simulations are consistent with those of F07.
For example, F07 lists $\alpha$ values for a $Re = 3125$ and $Rm =
12500$ model run with four different codes, including ZEUS.  These values
range from $\alpha=0.0091$ to $0.011$; we obtain $0.013$.  The increase
in turbulent energy levels with $\pr$ is demonstrated by a series of
simulations with the same $Rm$ and increasing viscosity.  For example,
for a constant $Rm \approx 12800$ (some of the simulations had $Rm =
12800$ while others had $Rm = 12500$; see F07),  $\pr$ values were
varied by factors of 2 from 1 to 16.  Sustained turbulence was seen
for $\pr \ge 4$ with $\alpha$ values increasing from $0.0091$
for $\pr = 4$ to $0.019$ and $0.044$ for $\pr =8$ and 16 respectively.
The $\at$ runs have $\alpha$ values of $0.013$, $0.026$, and $0.046$.
These data are plotted in Fig.~\ref{sz_rm12800}, which shows that the
increase in $\alpha$ with $\pr$ is nearly linear.

\begin{figure}
\plotfiddle{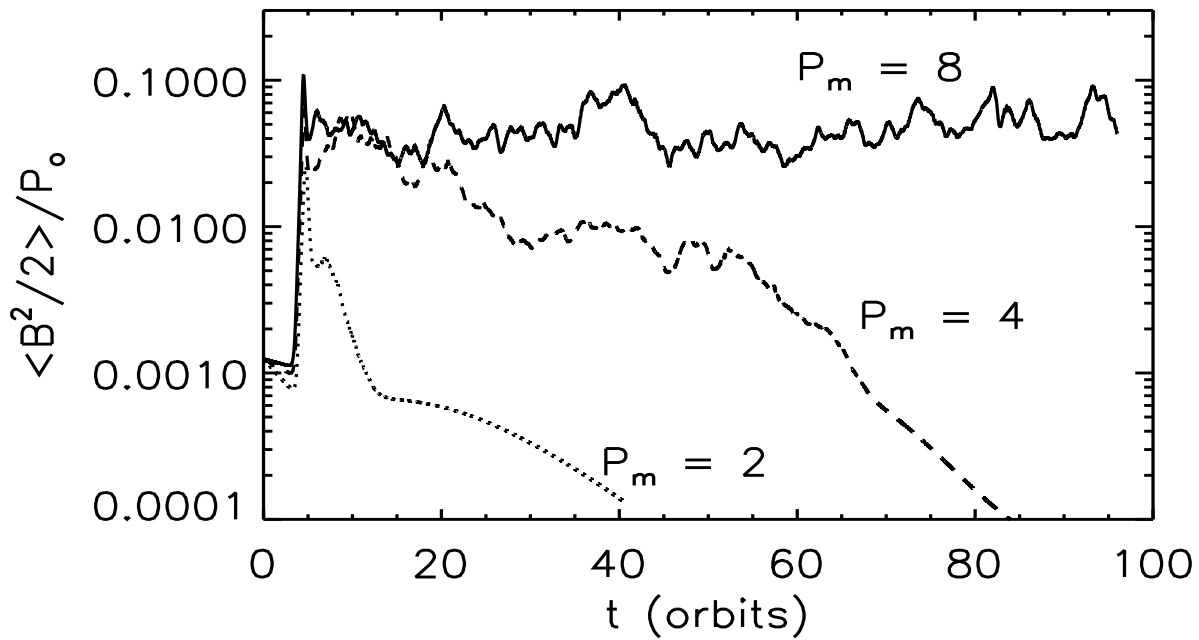}{0in}{0.}{470}{280}{0}{-100}
\vspace{-0.5in}
\caption{Time evolution of volume-averaged magnetic energy density
normalized by the gas pressure for the SZ runs with $Re = 1600$ and
varying $\pr$. The volume average is calculated over the entire simulation domain.
The solid line corresponds to $\pr$~=~8, the dashed line
corresponds to $\pr$~=~4, and the dotted line corresponds to $\pr$~=~2.
The turbulence decays for the lowest two $\pr$ values, with the $\pr$~=~4
case taking roughly 60 orbits to decay.
}
\label{sz_re1600}
\end{figure}

The largest differences between the $\at$ simulations and the ZEUS
simulations of F07 lie in the marginally turbulent cases.  For example,
we find decaying turbulence for $Re$~=~1600, $\pr$~=~4, whereas ZEUS
produces sustained turbulence for these parameters. Figure~\ref{sz_re1600}
shows the volume-averaged magnetic energy density normalized by the
gas pressure versus time for the three $\pr$ values at $Re$~=~1600.
The lowest $\pr$ simulation decays quite rapidly, whereas the $\pr$~=~4
case takes roughly 60 orbits to decay.  Differences in the numerical
properties of $\at$ and ZEUS might account for these results, given the
sensitivity to numerical factors as shown by zero net field simulations.
We also note that we use a slightly larger domain size in $y$ than in F07.
The boundary in parameter space between sustained turbulence and decay
is unlikely to be hard and fast, and detailed numerical surveys that
attempt to define that boundary are probably not warranted.  Some such
studies may, however, provide additional insights into the sensitivity
of the MRI turbulence to specific numerical factors.

\begin{deluxetable}{l|ccccc} 
\tabletypesize{\scriptsize}
\tablewidth{0pc}
\tablecaption{Zero Net Flux Simulations\label{tbl:sz_runs}}
\tablehead{
\colhead{Label}&
\colhead{$Re$}&
\colhead{$\pr$}&
\colhead{$Rm$}&
\colhead{Turbulence?}&
\colhead{$\alpha$}  }
\startdata
SZRe800Pm4  &  800 & 4 &  3200 & No  &    -  \\
SZRe800Pm8  &  800 & 8 &  6400 & Yes & 0.031 \\
SZRe800Pm16 &  800 &16 & 12800 & Yes & 0.046 \\
SZRe1600Pm2 & 1600 & 2 &  3200 & No  &     - \\
SZRe1600Pm4 & 1600 & 4 &  6400 & No  &    -  \\
SZRe1600Pm8 & 1600 & 8 &  12800& Yes & 0.026 \\
SZRe3125Pm1 & 3125 & 1 &  3125 & No  &    -  \\
SZRe3125Pm2 & 3125 & 2 &  6250 & No  &    -  \\
SZRe3125Pm4 & 3125 & 4 & 12500 & Yes & 0.013 \\
\enddata 
\end{deluxetable}

\section{Toroidal Field Simulations}
\label{toroidal_field}

To examine the effect of viscosity and resistivity on the MRI with a
net toroidal field, we have run a series of simulations initialized with
${\bm B} = \sqrt{2P/\beta}\hat{{\bm y}}$, where $\beta$~=~100, and with
varied $Re$ and $Rm$ values.  $Re$ ranges from
100 to 25600, and $\pr$ ranges from 0.25 to 16 (though, in some
simulations, we set either $\eta$ or $\nu$ equal to zero).  We will
consider the influence of the physical dissipation terms on two types of
problems:  the linear MRI growth regime, and fully nonlinear turbulence.

\subsection{The Linear Regime}
\label{linear_regime}

The linear nonaxisymmetric MRI was first examined by \cite{balbus92}.
For nonaxisymmetric modes, the MRI tends to be most robust in the
presence of a poloidal field.  However, even the purely toroidal field
case is unstable, athough, as emphasized by \cite{balbus92}, that case
is somewhat singular.  As always with the ideal MRI, the most unstable mode has
${\bf k \cdot v_A} \simeq \Omega$.  The linear analysis is complicated by
the background shear which causes radial wavenumbers to evolve with time.
Amplification of a given mode occurs  when the wavenumber ratio $k/k_z$
goes through a minimum as the radial wavenumber swings from leading
to trailing.  In general, the purely toroidal MRI favors high $k_z$
wavenumbers and small values of $k_y/k_z$, in contrast to the vertical
field MRI where the wavenumber $k_z$ of the most unstable mode is
determined by the $\alf$ speed.

\cite{papaloizou97} examined the toroidal field MRI with the addition
of resistivity.  They point out that because $k_x$ grows arbitrarily
large, all linear modes will eventually damp out in the presence of resistivity.
For small enough resistivities, however, there can be a period of growth
when $k_x \sim 0$.  For the MRI to become self-sustaining, this 
growth has to continue long enough for the perturbations to reach
nonlinear amplitudes.
Resistivity is also particularly important for the pure toroidal field MRI
because large $k_z$ is favored for mode growth.  
Equation (32) of \cite{papaloizou97} provides an approximate condition
for transient amplification of the MRI in the presence of resistivity.
For Keplerian shear and for modes where ${\bf k \cdot v_A} \sim \Omega$,
this reduces to the condition 
\begin{equation}\label{reslimit}
k_z^2 \eta \sim \Omega .
\end{equation}
In other words, there is no amplification of modes for which the diffusion
time is comparable to the orbital frequency.  Although viscosity was not
included in the analysis, one might expect it to be similarly influential.

Simulations of the linear growth of the MRI in the presence of resistivity
for a purely toroidal $\beta = 100$ initial field were first carried
out by \cite{fleming00} using a ZEUS code with an adiabatic equation of state. 
For this field strength, the critical MRI
wavelength in the azimuthal direction is $2 \pi v_A/\Omega \approx H$.
They found field decay for a $Rm = 2000$ simulation, but field growth
to turbulent saturation for $Rm = 5000$ and above.

In this section, we follow the growth of the
MRI in a shearing box with a purely toroidal field while including both
resistivity and viscosity.  The system is seeded within each grid zone
with random perturbations
in $\rho$ at amplitude $\delta \rho/\rho$~=~0.01. The simulations
were run at two resolutions, $N_x$~=~64, $N_y$~=~100, $N_z$~=~64 and
$N_x$~=~128, $N_y$~=~200, $N_z$~=~128 and are labelled YL for y-field,
linear regime.  In this standard set of simulations, the range of $Re$ examined runs from 800 to 25600,
and the range of $Rm$ is from 400 to 102400.
Table \ref{tbl:yl_runs} lists these simulations. The last two columns
state whether or not MRI growth is observed for the $N_x = 64$ and $N_x =
128$ resolutions, respectively.  A dash in either of these columns means
that the simulation was not run at that particular resolution.  MRI growth
is defined by the evolution of the volume-averaged magnetic and kinetic
energy components.  A simulation is considered to have zero growth if
after 20--40 orbits, the various energy components are either decaying
or constant in time without any indication of exponential increase. 
Growth to saturation is observed in cases when $Re$ and $Rm$ are at 6400 and
above.

Clearly, a sufficiently large viscosity or resistivity can
inhibit growth.  But what about the very high or very low $\pr$ limits?
To approach that question, we carried out simulations where only $\nu$
or $\eta$ was nonzero.  These experiments were done at the $N_x = 64$
resolution.  In our first experiments, we set $\eta$ to zero and $Re$
to 100 and 800. The $Re = 800$ run showed growth to saturation, but the
$Re = 100$ case had no growth.  Next we set $\nu$ to zero and $Rm$
to 800 and 1600. The lower resistivity ($Rm =
1600$) grew to saturation, whereas the higher resistivity ($Rm = 800$)
did not.  Although the existence of a critical $Rm$ value is consistent
with the results of \cite{fleming00}, the value of $Rm$ at which growth
is prevented is smaller here than what they found.  We note that there remains
unavoidable numerical dissipation associated with grid scale effects,
which will make the value of a critical $Rm$ obtained through simulations
somewhat dependent on algorithm and resolution.

The effect of numerical resolution is not necessarily obvious.
Consider model YLRe3200Pm2, which has $Re = 3200$ and $Rm=6400$, and model
YLRe6400Pm0.5, which has these values reversed.  In both cases,
the $N_x = 64$ simulations show growth but the $N_x = 128$ models
do not.  One difference between the two resolutions is in the initial
perturbations.  While the density perturbations have the same amplitude
in both resolutions, the higher resolution initial density is given
power at smaller scales because the perturbations are applied to each
grid zone.  This leads to a smaller amplitude for each Fourier mode.
Does this account for the difference seen in these two resolutions?
To investigate this, we ran both $N_x = 64$ versions of YLRe3200Pm2
and YLRe6400Pm0.5 with initial perturbations of amplitude $\delta
\rho/\rho$~=~0.005 and $\delta \rho/\rho$~=~0.001.  Note that these
amplitudes lead to comparable ($\delta \rho/\rho$~=~0.005) or smaller
($\delta \rho/\rho$~=~0.001) amplitude modes in Fourier space compared to
the $\delta \rho/\rho$~=~0.01 initialized modes at the higher resolution.
Neither of the smaller amplitude YLRe3200Pm2 simulations showed any growth
(as of 20-30 orbits), but both YLRe6400Pm0.5 simulations showed growth to saturation.

From these experiments it seems that the effects of viscosity and
resistivity are comparable and that the transition region between
decay and growth to turbulence lies between Reynolds numbers of
3200 and 6400 for $\pr$ near unity.  This corresponds to a critical
vertical wavelength, defined in terms of equation~(\ref{reslimit}),
of $\lambda_c/H \sim 2\pi/{Rm}^{1/2} = 0.111$ and 0.079, respectively.
As viscosity (resistivity) is increased, MRI growth can be achieved
by decreasing the resistivity (viscosity). This trend only works up to
certain limits; if either the viscosity or resistivity is large enough,
MRI growth is completely quenched, independent of the value of the other
dissipation term.

\begin{deluxetable}{l|cccccc} 
\tabletypesize{\scriptsize}
\tablewidth{0pc}
\tablecaption{Toroidal Field Simulations Initialized from Linear Perturbations\label{tbl:yl_runs}}
\tablehead{
\colhead{Label}&
\colhead{$Re$}&
\colhead{$\pr$}&
\colhead{$Rm$}&
\colhead{$\Lambda$} &
\colhead{$N_x = 64$}&
\colhead{$N_x = 128$} }
\startdata
YLRe800Pm0.5  &  800 &0.5 &   400 &  8 &  No & -  \\
YLRe800Pm1    &  800 &  1 &   800 & 16 &  No & -  \\
YLRe800Pm2    &  800 &  2 &  1600 & 32 &  No & -  \\
YLRe800Pm4    &  800 &  4 &  3200 & 64 &  No & -  \\
YLRe800Pm8    &  800 &  8 &  6400 &128 &  No & -  \\
YLRe1600Pm0.5 & 1600 &0.5 &   800 & 16 &  No & -  \\
YLRe1600Pm1   & 1600 &  1 &  1600 & 32 &  No & -  \\
YLRe1600Pm2   & 1600 &  2 &  3200 & 64 &  No & -  \\
YLRe1600Pm4   & 1600 &  4 &  6400 &128 &  No & -  \\
YLRe1600Pm8   & 1600 &  8 & 12800 &256 &  No & -  \\
YLRe3200Pm0.5 & 3200 &0.5 &  1600 & 32 &  No & No \\
YLRe3200Pm1   & 3200 &  1 &  3200 & 64 &  No & No \\
YLRe3200Pm2   & 3200 &  2 &  6400 &128 & Yes & No \\
YLRe3200Pm4   & 3200 &  4 & 12800 &256 & - & Yes \\
YLRe6400Pm0.5 & 6400 &0.5 &  3200 & 64 & Yes & No  \\
YLRe6400Pm1   & 6400 &  1 &  6400 &128 & Yes & Yes \\
YLRe6400Pm2   & 6400 &  2 & 12800 &256 & Yes & Yes \\
YLRe6400Pm4   & 6400 &  4 & 25600 &512 & Yes & Yes \\
YLRe12800Pm0.5&12800 &0.5 &  6400 &128 & Yes & Yes \\
YLRe12800Pm1  &12800 &  1 & 12800 &256 & Yes & Yes \\
YLRe12800Pm2  &12800 &  2 & 25600 &512 & Yes & Yes \\
YLRe12800Pm4  &12800 &  4 & 51200 &1024& Yes & Yes \\
YLRe25600Pm0.5&25600 &0.5 & 12800 &256 & Yes & Yes \\
YLRe25600Pm1  &25600 &  1 & 25600 &512 & Yes & Yes \\
YLRe25600Pm2  &25600 &  2 & 51200 &1024& Yes & Yes \\
YLRe25600Pm4  &25600 &  4 &102400 &2048& Yes & Yes \\
\enddata 
\end{deluxetable}

\subsection{The Nonlinear Regime}
\label{nonlinear_regime}

Of potentially greater interest than the linear MRI regime is the effect
of viscosity and resistivity on fully developed MRI-driven turbulence.
To study this nonlinear regime, we begin with model YLRe25600Pm4, a
simulation with $Re = 25600$ and $\pr = 4$ at $N_x$~=~128, $N_y$~=~200,
$N_z$~=~128 (Table~\ref{tbl:yl_runs}) that was run to 59 orbits in time.
The MRI grows and the flow becomes fully turbulent.  Averaging from
$t=15$ to 59 orbits gives a stress value of $\alpha = 0.05$.  We use
this simulation at $t=36$ orbits to initialize a series of simulations
with different values of $Re$ and $Rm$.  These runs are labelled YN
for y-field, nonlinear regime, and they are all run to 200 orbits,
except for simulation YNRe12800Pm0.25, which was run to 100 orbits.
All the YN simulations are listed in Table~\ref{tbl:yn_runs}.

When evolving onward from orbit 36 with modified dissipation terms,
a simulation shows a rapid readjustment followed by either sustained
turbulence at a new amplitude or decay to smooth flow, depending on the
new values of $Re$ and $Rm$.  The column labeled ``Turbulence?"
in Table~\ref{tbl:yn_runs} states whether or not the given simulation has sustained turbulence.
Note that for $Rm \gtrsim 1600$, the turbulence is sustained except
for the relatively viscous $Re=400$ model.  This critical $Rm$ value
is below the critical value obtained above for sustained growth in
the linear regime when the resistivity and viscosity are comparable
but near the critical $Rm$ value in the linear regime in the absence
of explicit viscosity.  For simulations where turbulence is sustained,
the column labeled ``$\alpha$" gives the time- and volume-averaged dimensionless stress,
where the time average is calculated onward from orbit 50.

\begin{figure}
\plotfiddle{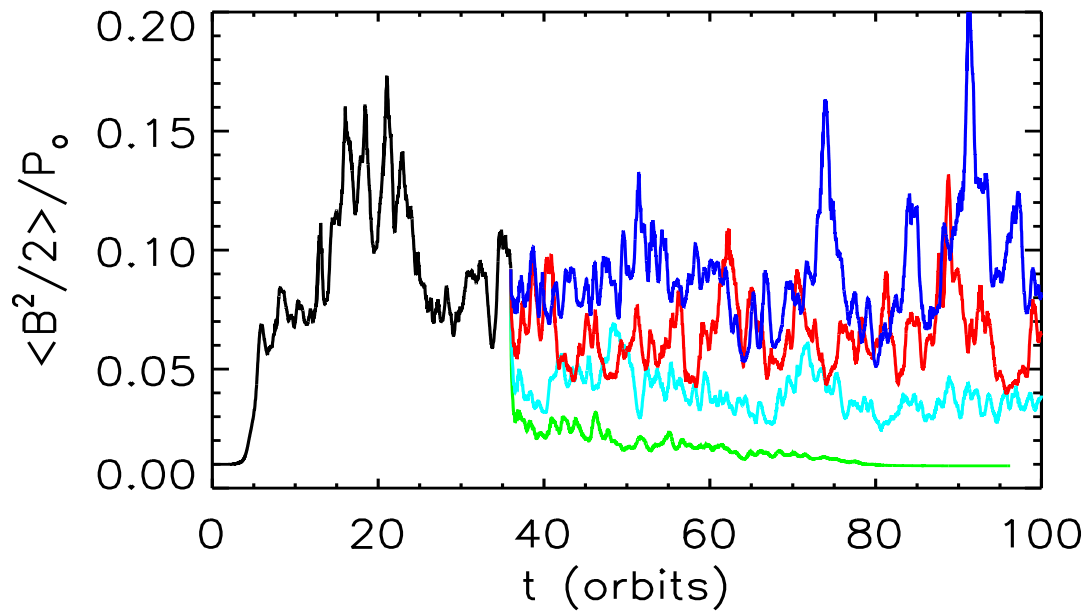}{0in}{0.}{450}{270}{0}{-100}
\vspace{-0.2in}
\caption{Time evolution of volume-averaged magnetic energy density
normalized by the gas pressure for the YN runs with $Re = 25600$ (black
curve) and $Re = 1600$ (colored curves).  The volume average is calculated over the entire simulation domain.
The colors indicate $\pr$;
green corresponds to $Rm = 800$ ($\pr = 0.5$), light blue to $Rm = 1600$ ($\pr =
1$), red to $Rm = 3200$ ($\pr = 2$), and dark blue to
$Rm = 6400$ ($\pr = 4$).  Increasing $Rm$ ($\pr$) leads to enhanced turbulence.
\label{re1600_be}}
\end{figure}

The column labeled ``$\langle\langle \Lambda\rangle\rangle$" gives a time- and volume-averaged $\Lambda$ value
in the final state of each simulation.  Unlike $Rm$, $\Lambda$ will
change with the evolving  magnetic field strength.
Beginning with equation~(\ref{rm_els_relate}), we write

\begin{equation}
\label{nonlinear_beta}
\beta = \frac{2 \cs^2 \langle \rho \rangle}{\langle B^2\rangle}
\end{equation} 

\noindent
to give
\begin{equation}
\label{nonlinear_els}
\langle \Lambda \rangle = 
\frac{Rm}{\cs^2} \frac{\langle B^2\rangle}{\langle \rho \rangle},
\end{equation}

\noindent where the angled brackets denote a volume average.
One could also volume-average the square of the $\alf$ speed
in the calculation of $\beta$ instead of averaging $B^2$ and $\rho$
separately (e.g., $\beta = 2 \cs^2/\langle \va^2 \rangle$).  We have
calculated $\langle \Lambda \rangle$ using both types of averages
for several frames in the saturated state of a few simulations.
We have found at most a factor of 2 difference between the different
calculations.  Since $\langle B^2 \rangle$ varies by a similar factor
in the saturated state (see Fig.~\ref{re1600_be}), this factor of 2
difference is within the uncertainty of $\Lambda$ at any
given time.  The time-average of the volume-averaged Elsasser number,
$\langle \langle \Lambda \rangle \rangle$, as given in the table, is
calculated from orbit 50 until the end of the simulation. For the decayed
turbulence simulations in which the turbulence has not fully decayed by orbit 50,
the time average is calculated onward from a point at which the volume-averaged
magnetic energy is constant in time.  Note that for these 
decayed turbulence simulations, $\langle\langle\Lambda\rangle\rangle$ should equal the $\beta = 100$
 value associated with the net toroidal field, as given in Table~\ref{tbl:yl_runs}.
 However, because of the evolution of the net
 $B_y$ (see \S~\ref{method}), the value of $\langle\langle\Lambda\rangle\rangle$ after the turbulence
 has decayed will be slightly different than the $\beta = 100$ value.  

Since the magnetic field varies within the domain, the local value of
$\Lambda$ can also vary from the overall average.  Histograms showing
the number of grid zones with $\va^2$ of a certain value reveal that the
percentage of grid zones that have $\Lambda < 1$ is at most $\sim0.01$\%.
For the sustained turbulence models, $\langle \langle \Lambda \rangle
\rangle$ is typically on the order of 100-1000; the smallest value for
a run with sustained turbulence is 106, and the largest value associated
with a run that decays is 30.

\begin{deluxetable}{l|ccccccc} 
\tabletypesize{\scriptsize}
\tablewidth{0pc}
\tablecaption{Toroidal Field Simulations Initialized from Nonlinear Turbulence\label{tbl:yn_runs}}
\tablehead{
\colhead{Label}&
\colhead{$Re$}&
\colhead{$\pr$}&
\colhead{$Rm$}&
\colhead{Turbulence?}&
\colhead{$\alpha$}& 
\colhead{$\langle\langle\Lambda\rangle\rangle$}& 
\colhead{$\langle\langle\Lambda_z\rangle\rangle$}    }
\startdata
YNRe400Pm0.5    & 400   & 0.5 & 200  & No &    -  &  4 & - \\
YNRe400Pm1      & 400   & 1   & 400  & No &    -  &  8 & - \\
YNRe400Pm2      & 400   & 2   & 800  & No &    -  &  15 & - \\
YNRe400Pm4      & 400   & 4   & 1600 & No &    -  &  30 & - \\
YNRe400Pm8      & 400   & 8   & 3200 & Yes& 0.043 &  614 & 16.8 \\
YNRe400Pm16     & 400   & 16  & 6400 & Yes& 0.068 &  1983 & 58.2 \\
YNRe800Pm0.25   & 800   & 0.25& 200  & No &    -  &   4 & - \\
YNRe800Pm0.5    & 800   & 0.5 & 400  & No &    -  &   8 & - \\
YNRe800Pm1      & 800   & 1   & 800  & No &    -  &  15 & - \\
YNRe800Pm2      & 800   & 2   & 1600 & Yes& 0.019 &  137 &  3.87 \\
YNRe800Pm4      & 800   & 4   & 3200 & Yes& 0.038 &  495 & 18.0 \\
YNRe800Pm8      & 800   & 8   & 6400 & Yes& 0.054 &  1413 & 56.2 \\
YNRe1600Pm0.5   & 1600  & 0.5 & 800  & No &    -  &   15 & - \\
YNRe1600Pm1     & 1600  & 1   & 1600 & Yes& 0.018 &  120 & 4.45 \\
YNRe1600Pm2     & 1600  & 2   & 3200 & Yes& 0.033 &  403 & 18.6 \\
YNRe1600Pm4     & 1600  & 4   & 6400 & Yes& 0.044 &  1120 & 52.6 \\
YNRe3200Pm0.5   & 3200  & 0.5 & 1600 & Yes& 0.016 &  106 & 4.53 \\
YNRe3200Pm1     & 3200  & 1   & 3200 & Yes& 0.025 &  314 & 16.4 \\
YNRe3200Pm2     & 3200  & 2   & 6400 & Yes& 0.035 &  860 & 47.4 \\
YNRe3200Pm4     & 3200  & 4   & 12800& Yes& 0.043 &  2170 & 127 \\
YNRe6400Pm0.5   & 6400  & 0.5 & 3200 & Yes& 0.021 &  263 & 14.9 \\
YNRe6400Pm1     & 6400  & 1   & 6400 & Yes& 0.031 &  748 & 45.2 \\
YNRe6400Pm2     & 6400  & 2   & 12800& Yes& 0.038 &  1880 & 118 \\
YNRe12800Pm0.25 & 12800 & 0.25& 3200 & Yes& 0.021 &  262 &  15.8 \\
\enddata 
\end{deluxetable}

The behavior of the MRI is often characterized by the vertical component of the $\alf$ speed, and
as such, we have also calculated the Elsasser number using only the vertical component
of the magnetic field,

\begin{equation}
\label{els_z}
\langle \Lambda_z \rangle = 
\frac{Rm}{\cs^2} \frac{\langle B_z^2\rangle}{\langle \rho \rangle},
\end{equation}

\noindent
where the angled brackets denote a volume average.  We have calculated the time average
of this number, $\langle\langle \Lambda_z\rangle\rangle$, onward
from orbit 50 for all the sustained turbulence YN simulations.  This number is displayed
 in the last column of Table~\ref{tbl:yn_runs}.  The decayed turbulence simulations have 
 $B_z$ approaching zero, and we do not calculate a vertical Elsasser number for these.  Again,
 we calculated the vertical Elsasser number both by averaging $B_z^2$ and $\rho$ separately as well as
 by averaging the ratio $B_z^2/\rho$.  We compared the two calculations for several frames and found
 at most a factor of 1.3 difference between them.  

The $\langle\langle \Lambda_z\rangle\rangle$ values for the runs that have $Rm$ closest
to the critical value are on the order unity, with the smallest value being 3.87.  
As touched upon by \cite{fleming00}, growth of the vertical field MRI is largely suppressed
for $v_{{\rm A}z}^2/(\eta \Omega) \lesssim 1$ (i.e., for vertical Elsasser numbers less than unity).
That we find $\langle\langle \Lambda_z\rangle\rangle \sim 1$ close to the ``decayed turbulence" regime
may suggest that the vertical field MRI plays an important role in the sustained nonlinear turbulence of
these toroidal field simulations.  One trend to note from these data is that the ratio of $\langle\langle \Lambda_z\rangle\rangle$
to $\langle\langle \Lambda\rangle\rangle$ increases with both decreasing $\nu$ and decreasing $\eta$;  the
vertical magnetic energy becomes a larger fraction of the total magnetic energy as either dissipation term
is reduced.  

The evolution of the magnetic energy in a typical set of simulations
is shown in Fig.~\ref{re1600_be}.  For these runs, $Re=1600$ and $Rm$
varies by factors of two from $Rm=800$ to 6400.  The black line shows
the initial evolution of YLRe25600Pm4, whose state at 36 orbits serves
as the initial condition.  It is clear that decreasing the resistivity
(increasing the $\pr$ number) enhances the saturation level, and for a
large enough resistivity, the turbulence decays.

\begin{figure}
\plotfiddle{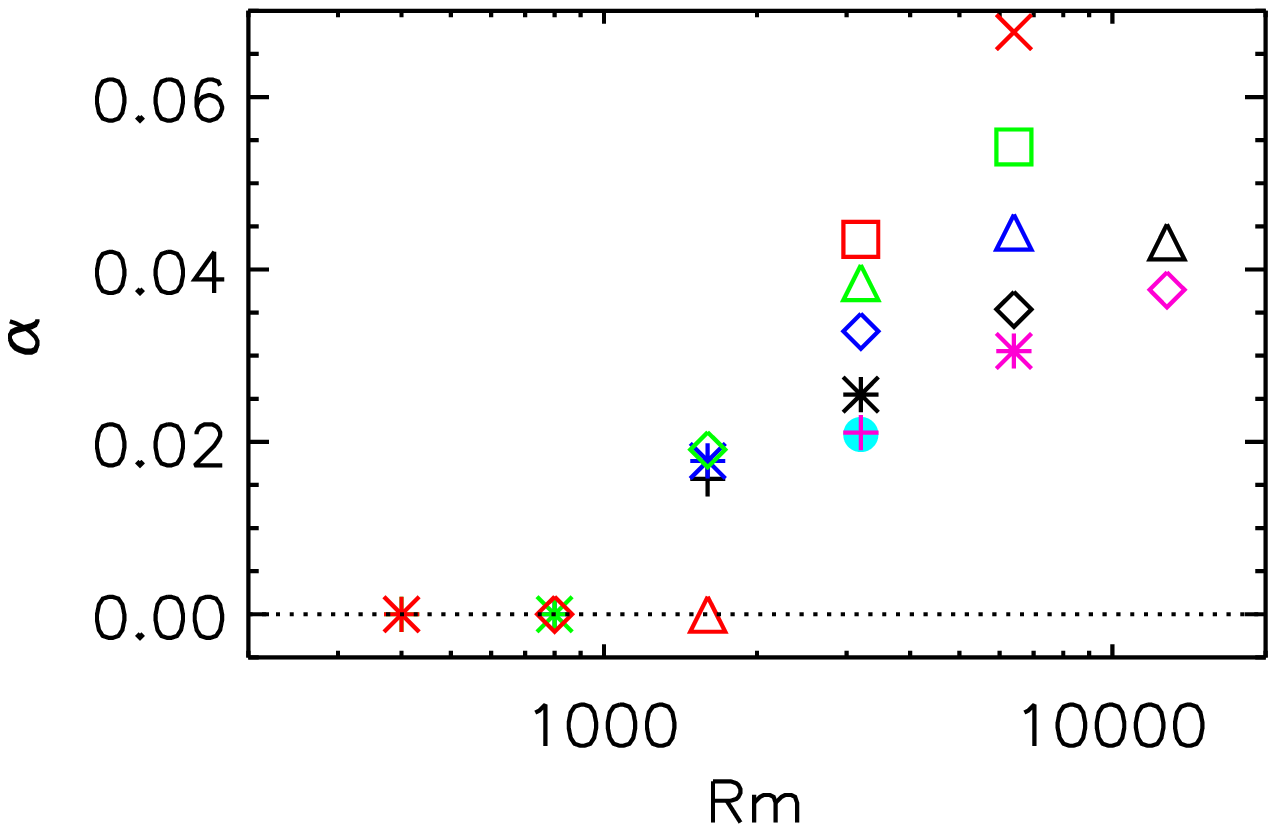}{0in}{0.}{400}{270}{10}{-100}
\vspace{-0.2in}
\caption{Time- and volume-averaged stress parameter $\alpha$ as a
function of $Rm$ in the YN simulations; $\alpha \equiv \langle\langle
\rho v_x\delta v_y - B_xB_y \rangle\rangle/P_o$.  The time average runs
from 50 orbits onward, and the volume average is calculated over the
 entire simulation domain. The colors correspond to $Re$ values, and the
symbols correspond to $\pr$ values.  Red symbols are $Re = 400$, green
$Re = 800$, dark blue $Re = 1600$, black $Re = 3200$, pink $Re = 6400$,
and light blue are $Re = 12800$.  Circles are $\pr = 0.25$, crosses $\pr
= 0.5$, asterisks $\pr = 1$, diamonds $\pr = 2$, triangles $\pr = 4$,
squares $\pr = 8$, and X's are $\pr = 16$. Note that some of the decayed
turbulence ($\alpha = 0$) simulations are not plotted for clarity.
Increasing $Rm$ results in larger $\alpha$ values, and for
$Rm$ less than 800--1600, the turbulence decays. 
}
\label{alpha_rm}
\end{figure}

To quantify the dependence of the saturation amplitude on the dissipation
coefficients, we plot the $\alpha$ values for the ensemble of simulations as a
function of $Re$, $Rm$ and $\pr$.  Figure~\ref{alpha_rm} shows $\alpha$
versus $Rm$; the color indicates $Re$ value, and the symbols correspond
to the $\pr$ value.  The simulations with $\alpha = 0$ are those where
the turbulence decayed away, which include all simulations with $Rm \le
800$ and the $Re = 400$, $Rm = 1600$ simulation.  Overall there is a
general trend of increasing $\alpha$ value with decreasing
resistivity.

The dependence of $\alpha$ on $Re$ is shown in Fig.~\ref{alpha_re}.
Here the color indicates the $Rm$ value, whereas $\pr$ is again
represented by a symbol.  Evidently, if the resistivity is low enough,
increasing the viscosity will increase the $\alpha$ values.  However,
consider the YN simulations with $Rm = 1600$.  These simulations suggest
that if the resistivity is close to some critical value, increasing the
viscosity will cause the turbulence to decay.  Another feature of note is
that as $Re$ increases, the range of $\alpha$ for different $Rm$ values
becomes smaller, and $\alpha$  appears to converge to $\sim 0.02-0.04$ for
all $Rm$.   This could indicate that as $\nu$ and $\eta$ decrease, their
influence on the turbulence level might decrease.  However, for large
values of $Re$ or $Rm$, the dissipation lengthscales are under-resolved,
and higher resolution is needed to test this possibility.

\begin{figure}
\plotfiddle{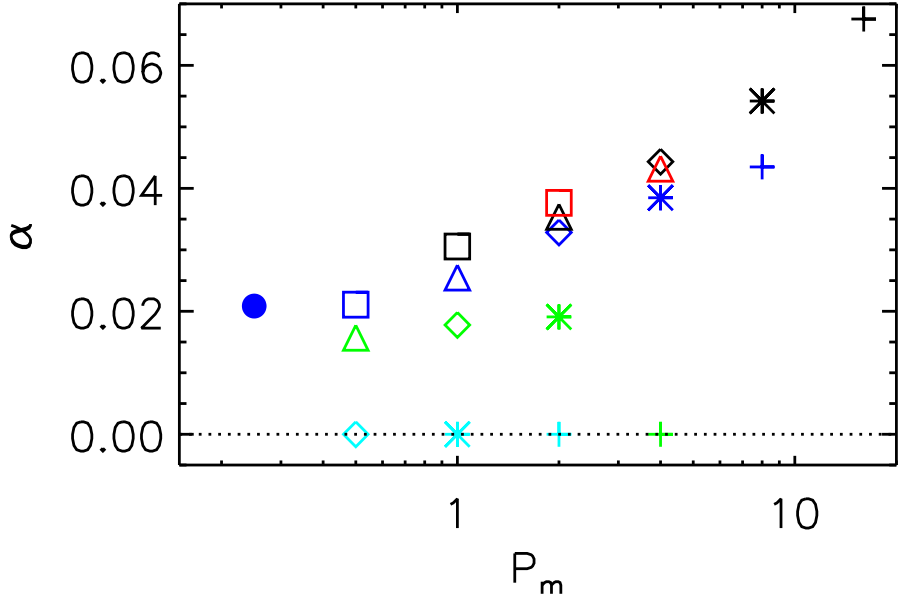}{0in}{0.}{400}{270}{10}{-100}
\vspace{-0.2in}
\caption{Time- and volume-averaged stress parameter $\alpha$ as a function
of $\pr$; $\alpha \equiv \langle\langle
\rho v_x\delta v_y - B_xB_y \rangle\rangle/P_o$.  The time average runs
from 50 orbits onward, and the volume average is calculated over the
 entire simulation domain.   The colors correspond to $Rm$ values, and the symbols to $Re$
values.  Light blue symbols are $Rm = 800$, green $Rm = 1600$, dark blue $Rm =
3200$, black $Rm = 6400$, and red are $Rm = 12800$.  
Crosses are $Re = 400$, asterisks 
$Re = 800$, diamonds $Re = 1600$, triangles $Re = 3200$, squares
$Re = 6400$, and circles are $Re = 12800$. Note that some of the decayed
turbulence ($\alpha = 0$) simulations are not plotted for clarity.  The 
average stress increases with increasing $\pr$.
\label{alpha_pm}}
\end{figure}

We plot the dependence of $\alpha$ on $\pr$ in
Fig.~\ref{alpha_pm}.  In this figure, the colors represent varying $Rm$
while the symbols denote different $Re$ values.  The clearest trend is
that if $Rm$ is large enough to sustain turbulence, increasing $\pr$
leads to larger $\alpha$ values.  Note that turbulence can be sustained
even for $\pr$ less than unity, if $Rm$ is large enough.  At constant
$Rm$, we find that $\alpha \propto Re^{\delta_1}$ with $\delta_1$ ranging
from -0.1 to -0.3 (calculated by a linear fit to the data in log space
for non-decayed turbulence simulations only).  At constant $Re$ value,
we find $\alpha \propto Rm^{\delta_2}$ with $\delta_2$ in the range
0.4-0.8 and $\delta_2$ generally decreasing with increasing $Re$.

These results naturally lead to the question of why increasing
$\nu$ or decreasing $\eta$ causes an increase in turbulence.
Magnetic reconnection and dissipation of field lines, either due to
an explicit resistivity or to grid-scale effects, presumably play
the primary role in limiting the amplitude of the MHD turbulence.
\cite{balbus98} hypothesized that increased viscosity would inhibit
reconnection by preventing velocity motions that would bring field
together on small scales.  When $\pr > 1$, the viscous length is greater
than the resistive one, and magnetic field dissipation becomes less efficient,
leading to an increase in turbulent stress \cite[e.g.,][]{balbus08a}.
If this hypothesis is correct, there may also be a change in the dissipation of
kinetic and magnetic energy into heat.  To investigate this
possibility, we carry out an analysis of viscous and resistive heating for
several of the simulations.

Consider the volume-averaged kinetic and magnetic energy evolution
equations, equations (15) and (16) in \cite{simon09}, 

\begin{eqnarray}
\label{ke_eqn}
\kd & = & -\left\langle \del \cdot \left[{\bm v}\left(\frac{1}{2}\rho v^2+\frac{1}{2}B^2 + P + \rho \Phi \right) - {\bm B} ({\bm v} \cdot {\bm B})\right]\right\rangle \nonumber \\
& & + \left\langle\left(P + \frac{1}{2} B^2\right)\del \cdot {\bm v}\right\rangle-\left\langle{\bm B} \cdot ({\bm B} \cdot \del {\bm v})\right\rangle - \gd - \qk,
\end{eqnarray}

\noindent
and

\begin{equation}
\label{be_eqn}
\md = - \left\langle \del \cdot \left(\frac{1}{2}B^2{\bm v}
\right)\right\rangle - \left\langle \frac{1}{2} B^2 \del \cdot {\bm
v}\right\rangle + \left\langle{\bm B} \cdot ({\bm B} \cdot \del {\bm
v})\right\rangle- \qm .
\end{equation} 

\noindent
Here, $\kd$ and $\md$ are the time derivatives of the volume-averaged
kinetic and magnetic energies, respectively.  The time derivative of
the volume-averaged gravitational potential energy is given by $\gd$,
and $\qk$ and $\qm$ are the volume-averaged kinetic and magnetic energy
dissipation rates, respectively.  The gravitational potential is $\Phi = q \Omega^2
(\frac{L_x^2}{12}-x^2)$.

We determine $\qk$ and $\qm$ for select YN models by computing the time
average of each of the source terms in the energy evolution equations
using 200 data files equally spaced in time over 20 orbits.  We
assume $\gd$ is zero in the time-average; the analysis of \cite{simon09}
found $\gd$ is always negligibly small.  The time-derivatives, $\kd$
and $\md$, are calculated by differentiating the volume-averaged kinetic
and magnetic energy history data with respect to time and then sampling
these data to the times associated with the data files.  The dissipation
terms $\qk$ and $\qm$, which include both physical and numerical effects,
are the remainder after all the other terms are calculated.

Figure~\ref{qkqm} shows the ratio of the time-average $\langle \qk\rangle$
to $\langle\qm\rangle$ as a function of $\pr$ and $\alpha$ for select
YN runs.  The colors and symbols are the same as in Fig.~\ref{alpha_re}.
The time average is calculated from $t = 70-90$ orbits for YNRe400Pm16
(black X) and YNRe12800Pm0.25 (blue circle), $t = 110-130$ orbits for
YNRe800Pm2 (green diamond) and YNRe800Pm8 (black square), and $t =
110.6-130.6$ orbits for YNRe800Pm4 (blue triangle) and YNRe3200Pm4 (red
triangle).  The ratio of viscous to resistive heating generally increases
as either $\alpha$ or $\pr$ increases, although not monotonically.
The relative heating ratio is not simply proportional to $\pr$ as one
might naively expect.

The data suggest a general relationship between saturated
stress and $\langle\qk\rangle$/$\langle\qm\rangle$.  We know that
the stress level sets the {\it total} dissipation rate ($\qk + \qm$)
\cite[e.g.,][]{simon09}; stronger stresses extract more energy from the
background shear flow, and that turbulence is rapidly dissipated into
heat.  However, does stronger turbulence by itself change the heating ratio,
or is the change in the heating ratio mainly determined by changes in
$\pr$, which also increase the turbulence levels?  This question of
causality cannot be definitively answered from these data.

Further insight may come from examining the ratio of averaged 
Reynolds stress, $\langle\langle\rho v_x \delta v_y\rangle\rangle$,
to averaged Maxwell stress, $\langle\langle-B_x B_y\rangle\rangle$,
 as a function of $\alpha$; this is shown in Fig.~\ref{reymax_alpha}.  The colors and symbols are the same as in
Fig.~\ref{qkqm}.  The double brackets for the stresses denote time and
volume averages, where the time average is calculated over the same 20
orbit period as in Fig.~\ref{qkqm}.  There is a decrease
in the ratio of the Reynolds to Maxwell stress as the total stress increases.
These stresses are proportional to the perturbed magnetic and kinetic
energies at the largest scales, and if this continued down to the
dissipation scale, we might expect that the ratio $\langle\qk\rangle$/$\langle\qm\rangle$
 would behave similarly with $\alpha$.  In fact, the heating ratio
shows the opposite trend with $\alpha$, indicating that a transfer of
energy from magnetic to kinetic fluctuations must occur in the turbulent
cascade.

Past net toroidal field simulations without explicit dissipation
terms also find a trend for a decrease in the ratio of the Reynolds
to Maxwell stress with increasing $\alpha$ \cite[e.g.,][]{hawley95a}.
So this may be a general result independent of $\pr$.  The quantity
$\langle\qk\rangle$/$\langle\qm\rangle$ has not been extensively studied
in past shearing box simulations, but \cite{simon09} found a ratio of $\sim
0.6$ for a net vertical field model without explicit dissipation terms.

In summary, these observations are consistent with the hypothesis that
decreasing $\pr$ increases the efficiency of magnetic reconnection and
hence reduces the overall stress level.   However, a more in-depth study
would be required to better understand the full causal relationship
between the ratio of dissipation terms and the saturation levels.

\begin{figure}
\plotfiddle{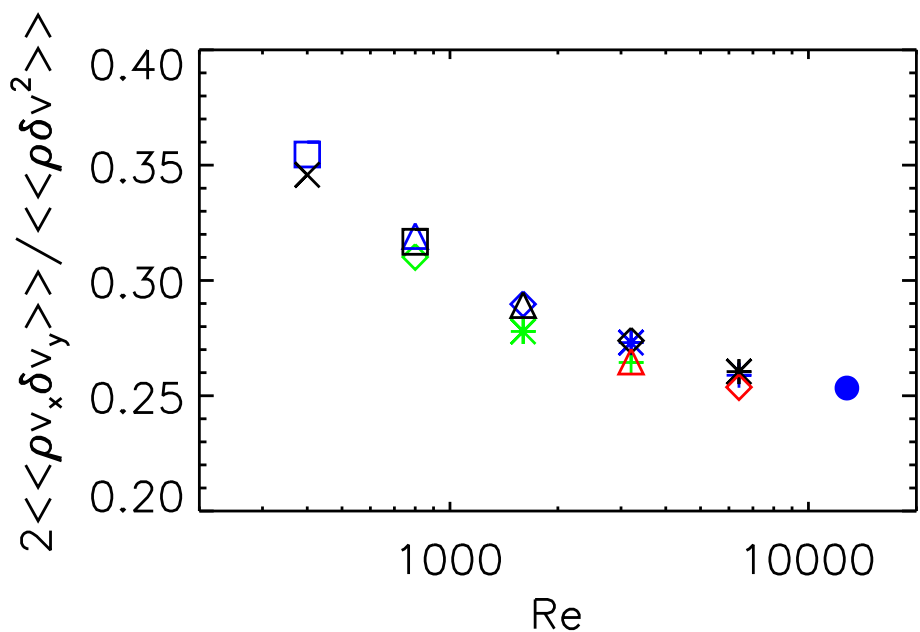}{0in}{0.}{400}{270}{10}{-100}
\caption{Ratio of Reynolds stress to perturbed kinetic energy as a function
of $Re$ in the sustained turbulence YN simulations.  Both the Reynolds stress
and the perturbed kinetic energy are time and volume averaged, with the time average calculated
from orbit 50 onward and the volume average calculated over the
entire simulation domain.   The colors correspond to $Rm$ values,
and the symbols correspond to $\pr$ values.  Green symbols are $Rm = 1600$, blue $Rm = 3200$, black $Rm =
6400$, and red are $Rm = 12800$. Circles are $\pr = 0.25$, crosses $\pr = 0.5$, asterisks $\pr =
1$, diamonds $\pr = 2$, triangles $\pr = 4$, squares $\pr = 8$, and
X's are $\pr = 16$.  The ratio of stress to energy increases with increasing viscosity but does
not change with resistivity.
\vspace{-0.2in} 
\label{reykp_re}}
\end{figure}

Finally, we note that the ratio of Reynolds stress to perturbed
kinetic energy increases with increasing $\nu$, as shown in Fig.~\ref{reykp_re}. 
There is no observed trend with $\eta$.  As
$\nu$ is increased, the fluid motions that are not being directly driven
by the MRI become increasingly damped.  The fluid motions
that are driven by the magnetic field in the form of Reynolds stress follow the behavior
of the Maxwell stress with $\nu$.  This is
also consistent with the hypothesis that increased $\nu$ leads to less efficient magnetic reconnection;
the kinetic fluctuations become damped relative to the driving via the MRI, making
it difficult to bring field lines close together for reconnection.   

Overall, resistivity seems to play a more fundamental role than viscosity
in these net toroidal field simulations.  There is a critical $Rm$ below
which turbulence decays or fails to grow from linear perturbations.
For a given resistivity near this critical value, a relatively low viscosity leads
to MRI growth (linear regime) or sustained turbulence (nonlinear regime).
A high viscosity can prevent growth (linear regime) or cause decay (nonlinear regime).  
Once the resistivity is sufficiently low to ensure MRI growth
to saturation and continued turbulence, the effect of viscosity changes
and higher viscosity gives larger $\alpha$ values.

\section{Discussion and Conclusions}
\label{conclusions}

In this study, we carried out a series of local, unstratified shearing box
simulations of the MRI with $\at$ including the effects of constant shear
viscosity and Ohmic resistivity.  The first simulations were initialized
with a zero net magnetic flux in the domain for comparison with the
results of F07.  The second
set of simulations are the first investigation of the impact of
both viscosity and resistivity on models with a net toroidal field.

For the values of viscosity and resistivity they studied, F07 found that
turbulence was sustained only above a critical $\pr$ value, specifically
when $\pr \gtrsim 1$.  There was evidence that this critical $\pr$ value
decreases as $Re$ increases (viscosity is reduced).  We repeated these
experiments and found that the saturation level of the MRI depends
strongly on both viscosity and resistivity, and for every $Re$, there
exists a critical $\pr$ value below which the turbulence dies out.  For
those simulations where turbulence was sustained, we found good agreement
between the $\at$ $\alpha$ values and those of F07.

Zero net field simulations are fundamentally different from net field
models because an imposed background field cannot be removed as a result
of the simulation evolution.  The net field remains unstable and can
drive fluid motion even during the fully nonlinear turbulence phase,
assuming that that field was unstable to begin with.  \cite{lesur07}
examined the effects of diffusion on models with a net $\beta =100$
vertical field in a $1\times4\times1$ shearing box using a pseudo-spectral
incompressible code.  They found a relation $\alpha \propto \pr^{\delta}$
with $\delta = 0.25$--0.5 for values of $\pr$ ranging from 0.12 to 8,
but they found no case where the turbulence died out completely for the
range of viscosities and resistivities studied.

Among net field models, the purely vertical field case is significantly
different from the purely toroidal field model, hence the need for the study
we have presented here.  For a vertical field, the linear MRI favors
wavenumbers $k_z \sim \Omega/\va$ and $k_x=k_y =0$. The purely toroidal
case favors $k_y \sim \Omega/\va$ with $k/k_z$ minimized.  Since $k_x$
is time dependent due to the background shear, a given mode undergoes a
finite period of amplification as $k_x$ swings from leading to trailing.
These properties suggest that purely toroidal field models might be
more sensitive to dissipation than the vertical field case.

In our numerical study of the linear growth regime of the toroidal MRI,
we have found that increasing either the viscosity or the resistivity
can prevent the growth of MRI modes.  As the viscosity (resistivity) increases,
the MRI needs a smaller resistivity (viscosity) in order to grow.  However, for
large enough values of either the viscosity or the resistivity, MRI growth
is suppressed, even in the absence of the other dissipation term.  Because of the
importance of small wavelength (large wavenumber) modes, the critical
$Rm$ values, below which growth is inhibited,
tends to be larger than what one would expect from an 
axisymmetric vertical field analysis, even in the absence of viscosity. 
Here, for comparable values of viscosity and resistivity, the critical 
$Rm$ value was around 3200--6400.

Because the linear toroidal field MRI is time dependent, turbulence
can only be sustained if nonlinear amplitudes are reached during the
growth phase.  Thus, the outcome of the linear MRI phase can be sensitive to
the initial amplitude of the perturbations in a simulation where the
viscous or resistive values are near the critical value.

In the nonlinear regime, we found that viscosity generally acts in
an opposite sense to that in the linear regime; increased viscosity
{\it enhances} angular momentum transport. Furthermore, increasing the
resistivity appears to decrease the saturation level, in agreement with
previous studies, and the critical $Rm$, below which the turbulence dies
is $\sim 800$--$1600$.  Near the critical $Rm$, however, increasing the
viscosity causes the turbulence to decay, a behavior more in line with
the linear regime.

In our simulations, as well as those of \cite{lesur07}, $\pr < 1$
did not necessarily quench the nonlinear turbulence or prevent growth
from linear perturbations. Resistivity or viscosity above a certain
level can stabilize the system against these perturbations, but if
both are sufficiently small, their ratio has no influence on MRI growth.
The {\it presence} of turbulence, however, is distinct from the saturation
level of that turbulence, and here $\pr$ can have a significant effect.
For both net toroidal and net vertical field simulations, $\alpha$ 
increases with increasing $\pr$ for the range of values studied.

\begin{figure}
\plotfiddle{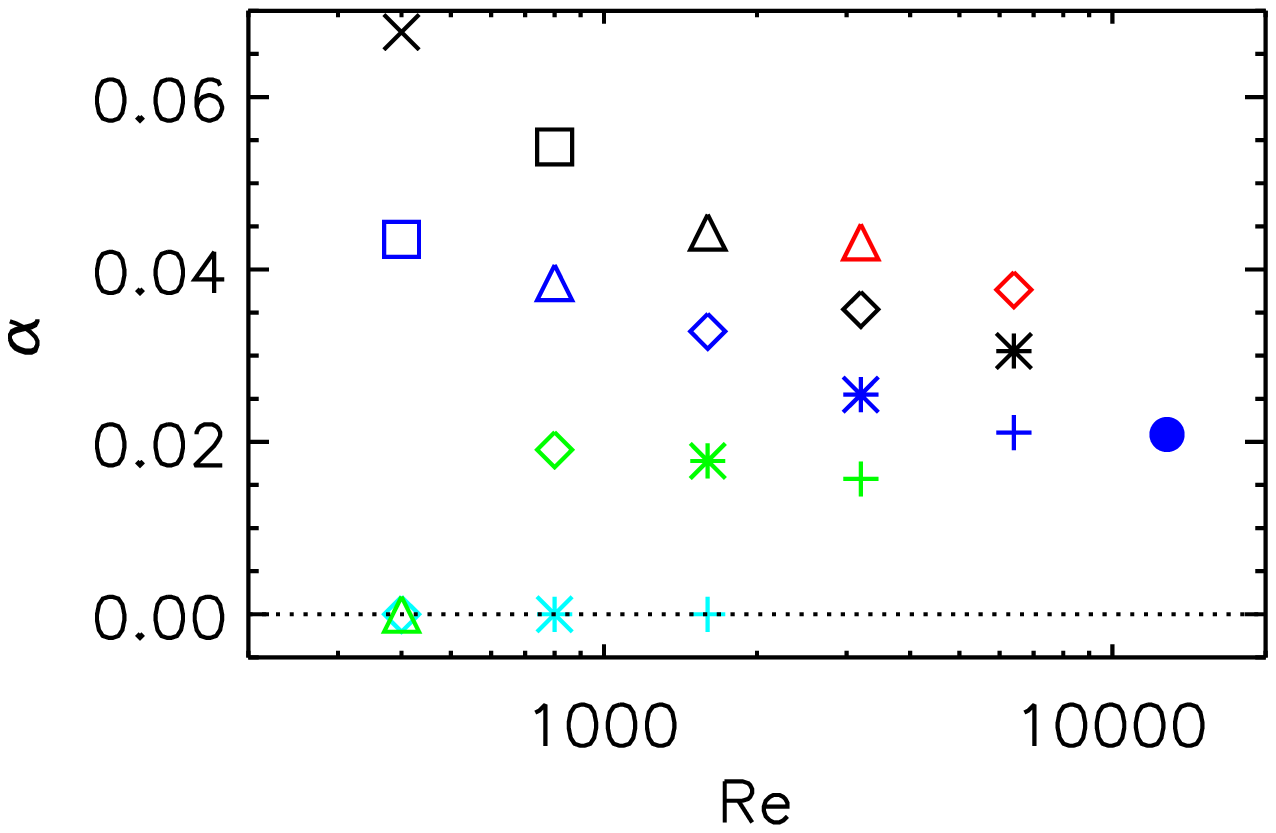}{0in}{0.}{400}{270}{10}{-100}
\vspace{-0.2in}
\caption{Time- and volume-averaged stress parameter $\alpha$ as a function
of $Re$ in the YN simulations; $\alpha \equiv \langle\langle
\rho v_x\delta v_y - B_xB_y \rangle\rangle/P_o$.  The time average runs
from 50 orbits onward, and the volume average is calculated over the
 entire simulation domain.   The colors correspond to $Rm$ values,
and the symbols correspond to $\pr$ values.  Light blue symbols are $Rm =
800$, green $Rm = 1600$, dark blue $Rm = 3200$, black $Rm =
6400$, and red are $Rm = 12800$. Circles are $\pr = 0.25$, crosses $\pr = 0.5$, asterisks $\pr =
1$, diamonds $\pr = 2$, triangles $\pr = 4$, squares $\pr = 8$, and
X's are $\pr = 16$. Note that some of the decayed
turbulence ($\alpha = 0$) simulations are not plotted for clarity.
Increasing $Re$ leads to decreasing $\alpha$ values.
\label{alpha_re}}
\end{figure}

\begin{figure}
\plotfiddle{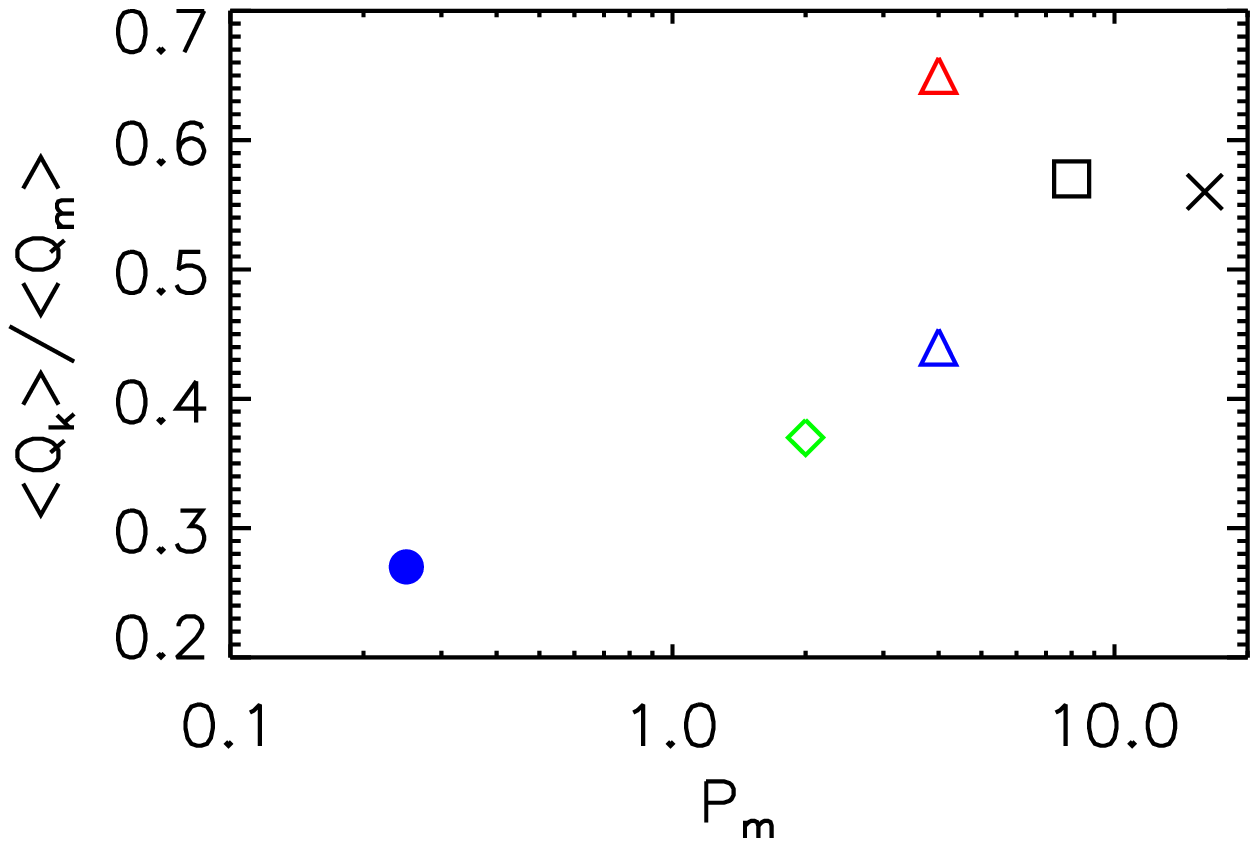}{0in}{0.}{300}{200}{90}{0}
\plotfiddle{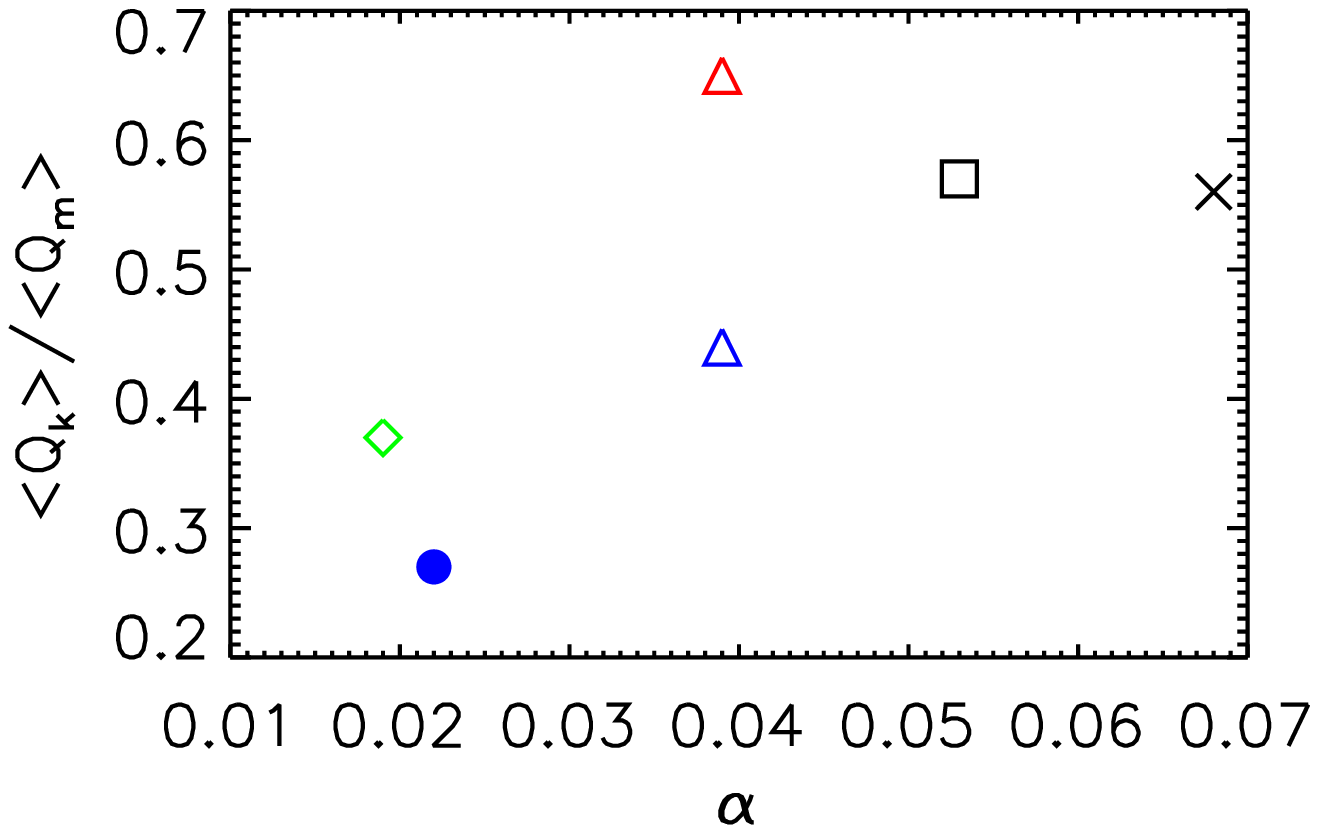}{0in}{0.}{300}{200}{90}{0}
\caption{Ratio of kinetic to magnetic energy dissipation rate
as a function of $\pr$ (top panel) and $\alpha$ (bottom panel) for select YN runs; $\alpha \equiv \langle\langle
\rho v_x\delta v_y - B_xB_y \rangle\rangle/P_o$.  The colors and symbols are the same as in Fig.~\ref{alpha_re}.
The kinetic and magnetic dissipation rates as well as $\alpha$ have been averaged in volume and time. 
The volume average is calculated over the entire simulation domain and the
 time average is calculated from $t = 70-90$ orbits for YNRe400Pm16
(black X) and YNRe12800Pm0.25 (blue circle), $t = 110-130$ orbits for
YNRe800Pm2 (green diamond) and YNRe800Pm8 (black square), and $t =
110.6-130.6$ orbits for YNRe800Pm4 (blue triangle) and YNRe3200Pm4 (red
triangle).  The ratio of viscous to resistive heating generally increases
as either $\alpha$ or $\pr$ increases, although not monotonically.
\vspace{-0.2in} 
\label{qkqm}}
\end{figure}

\begin{figure}
\plotfiddle{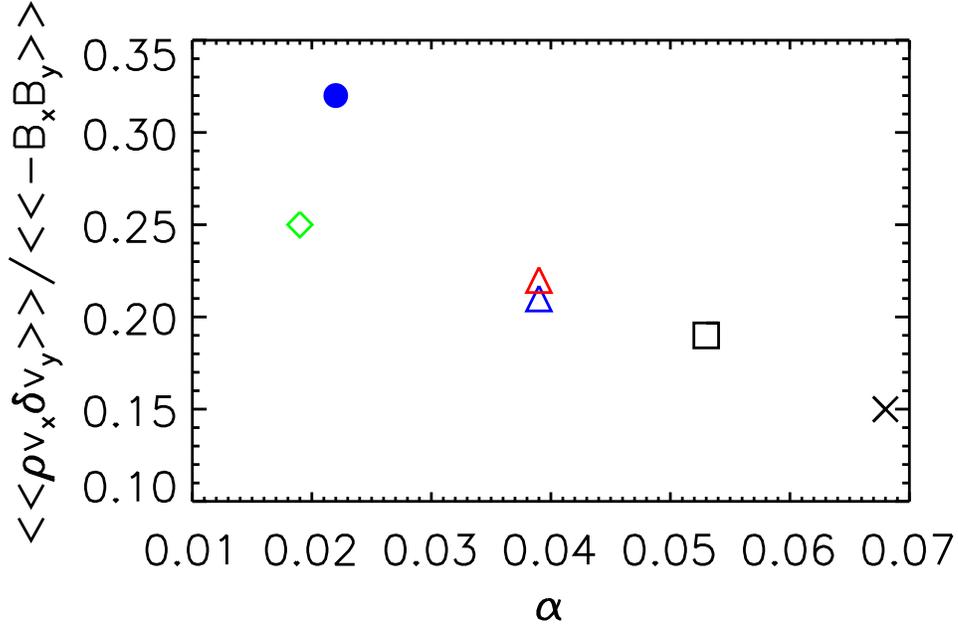}{0in}{0.}{400}{270}{10}{-100}
\caption{Ratio of Reynolds stress to Maxwell stress
as a function of $\alpha$ for select YN runs; $\alpha \equiv \langle\langle
\rho v_x\delta v_y - B_xB_y \rangle\rangle/P_o$. The colors and symbols are the same as in Fig.~\ref{alpha_re}.
The Maxwell and Reynolds stresses as well as
$\alpha$ have been averaged in volume and time.  
The volume average is calculated over the entire simulation domain and the
 time average is calculated from $t = 70-90$ orbits for YNRe400Pm16
(black X) and YNRe12800Pm0.25 (blue circle), $t = 110-130$ orbits for
YNRe800Pm2 (green diamond) and YNRe800Pm8 (black square), and $t =
110.6-130.6$ orbits for YNRe800Pm4 (blue triangle) and YNRe3200Pm4 (red
triangle).  The ratio of Reynolds to Maxwell stress generally decreases
with increasing $\alpha$.
\vspace{-0.2in} 
\label{reymax_alpha}}
\end{figure}

What do these results imply for the effect of resistivity and viscosity
on the MRI and on astrophysical systems? In principle, they could be
quite significant. In protostellar disks, the resistivity is thought to
be quite high near the midplane, leading to the existence of the dead
zone \cite[]{gammie96}. The $Rm$ values studied here could be applicable
to such systems. However, the implications for accretion disks with
small values of viscosity and resistivity (e.g., X-ray binary disks)
are less clear.  Because the range of $\alpha$ values we obtained decreases
with increasing $Re$ (Fig.~\ref{alpha_re}), it is possible that $\alpha$
may converge to a single value independent of $\pr$ as $Re$ and $Rm$
are increased.  If true, this would suggest that the dissipation
scales might have very little influence on the saturation level of
the MRI in astrophysical disks. This idea will need to be tested with
higher resolution simulations to ensure that the (small) viscous and
resistive scales are adequately resolved.  If, on the other hand, $\pr$
still has an influence on the turbulence even for very small values of
viscosity and resistivity, our results (taken together with those in
the literature) could be applicable to such disks.  The resistivity,
viscosity, and $\pr$ can vary quite substantially in these systems,
not only between different astrophysical objects, but also within a
given disk \cite[e.g.,][]{brandenburg05}.    \cite{balbus08a} analyze
a possible $\pr$ dependence on radius in X-ray binaries to
suggest that such a dependence could be at the core of spectral state
transitions in these systems.

The work to date is suggestive, but there remain several limitations
associated with these shearing box simulations.  
First, the simulations are unstratified; vertical gravity may also
play a role in establishing the overall turbulent state.  For example, \cite{davis09}
carried out a series of zero net field shearing box simulations with vertical gravity
and explicit dissipation and found that the turbulence does not decay as readily
as in the unstratified case.  Secondly,
all of the simulations to date have explored a relatively restricted
range of parameters.  Here, for example, we have examined only one
value for the toroidal field strength and one domain size.  Finally,
as touched upon above, the range of values for $Re$ and $Rm$ that have
been studied are somewhat restricted and certainly much smaller than
would be appropriate for many astrophysical disk systems.  While this limitation is
partially computational and can be improved upon with higher resolutions,
the question remains for astrophysical systems whether  viscous and
resistive processes that take place on relatively small lengthscales can
have a significant influence on macroscopic stress terms whose scales
are on order the pressure scale height in the disk.  But regardless of
the importance of resistivity and viscosity for astrophysical systems,
the values of $Re$ and $Rm$ are very important for setting $\alpha$ in
numerical simulations, much more so than many other shearing box parameters
(e.g., pressure) studied to date.  Without a more thorough understanding
of the role that dissipation terms play, quantitative predictions of $\alpha$
values from simulations will not be possible.

In summary, our experiments have explored the effect of changing viscosity
and resistivity on MRI simulations with a net toroidal field.  This work
serves to expand upon previous investigations of the impact of small-scale
dissipation.  While the direct applicability of studies such as this to
specific stress values within astrophysical systems remains uncertain,
it is likely that for the conceivable future, numerical simulations will
be our primary, if not only way to explore the nature of MRI-driven
turbulence.  A thorough understanding of MRI turbulence can only
be obtained with a complete understanding of the effects of diffusion,
both numerical and physical.

\acknowledgments

We thank Jim Stone, Steve Balbus, Jeremy Goodman, Pierre-Yves Longaretti, Martin Pessah, Xiaoyue Guan,
and Sebastien Fromang for useful discussions and suggestions regarding
this work. We also thank the anonymous referee whose comments and suggestions
improved this paper. This work was supported by NASA Headquarters under the
NASA Earth and Space Science Fellowship Program Grant NNX08AX06H, a
Virginia Space Grant Consortium fellowship, NASA grant NNX09AD14G, and NSF grant AST-0908869.
The simulations were run on the TeraGrid Ranger system at TACC, supported
by the National Science Foundation.

\clearpage

\clearpage

\clearpage

\clearpage

\clearpage

\clearpage

\clearpage

\clearpage

\clearpage

\clearpage

\end{document}